\documentstyle[11pt,a4,amssymb]{article}

\newcommand{\be}{\begin{equation}}
\newcommand{\ee}{\end{equation}}

\newcommand{\bea}{\begin{eqnarray}}
\newcommand{\eea}{\end{eqnarray}}

\begin{document}

\begin{titlepage}

\begin{flushright}
{\tt FTUV/95-17\\
     IFIC/95-17\\
     Imperial-TP/94-95/31\\
     hep-th/9505139}
\end{flushright}

\vfill

\begin{center}

{\bf{\Large
	Weyl Invariance and Black Hole Evaporation}\footnote{
		Work partially supported by the
		{\it Comisi\'on Interministerial de Ciencia y
		Tecnolog\'{\i}a} and {\it DGICYT}.}
}

\bigskip
\bigskip
\bigskip

	J. Navarro-Salas$^a$\footnote{\sc jnavarro@evalvx.ific.uv.es},
	M. Navarro$^b$\footnote{\sc m.navarro@ic.ac.uk}
	and C. F. Talavera$^a$\footnote{\sc talavera@evalvx.ific.uv.es}\\

\bigskip

\begin{center}
	$^a$ Departamento de F\'{\i}sica Te\'orica and\\
	IFIC, Centro Mixto Universidad de Valencia-CSIC.\\
	Facultad de F\'{\i}sica, Universidad de Valencia,\\
	Burjassot-46100, Valencia, Spain.\\

\bigskip

	$^b$ The Blackett Laboratory, Imperial College,\\
	London SW7 2BZ, United Kingdom.
\end{center}
\bigskip
\today
\bigskip

\end{center}

\begin{center}
{\bf Abstract}
\end{center}

We consider the semiclassical dynamics of CGHS black holes with a
Weyl-invariant effective action for conformal matter.
The trace anomaly of Polyakov effective action is converted into the
Virasoro anomaly thus leading to the same flux of Hawking radiation.
The covariance of semiclassical equations can be restored through a
non-local redefinition of the metric-dilaton fields.
The resulting theory turns out to be equivalent to the RST model.
This provides a mechanism to solve semiclassical equations of 2D dilaton
gravity coupled to conformal matter for classically soluble models.

\vfill

\end{titlepage}
\newpage

\section{Introduction}

	Since the pioneer work of Hawking \cite{Hawking}, the
formation and subsequent evaporation of a black hole has provided
an excellent scenario to study the interplay between gravity and
quantum mechanics.
In recent years it has been a revival of interest in the subject,
due to the emergence of simplified (two-dimensional) models
sharing basic features with the four-dimensional theory.
The model proposed by Callan, Giddings, Harvey and Strominger
\cite{CGHS} (CGHS-model), involving gravity coupled to a dilaton
and N massless scalar fields $f_i$, $i=1,\dots,N$,
describes, at the classical level,
the formation of a black hole by incoming conformal matter.
Hawking radiation in the classical background geometry can be
computed from the trace anomaly of the matter fields
\cite{Christensen}
\be
	< {{T^f}^\alpha}_\alpha > = {N \over 24} R \> ,
\label{i}
\ee
and back-reaction can be incorporated by adding to the
classical action the Polyakov effective action
\cite{Polyakov}
\be
	S_P = - {N \over 96\pi} \int d^2 x \sqrt{-g} R \square^{-1} R
\>.
\label{ii}
\ee
The new equations of motion have the quantum stress tensor of
matter as the source for the classical gravity dilaton fields.
Although the semiclassical equations have not been solved in closed
form, a special modification of the model \cite{RST} allows to
construct exact solutions and, therefore, to study the evolution
of a quantum black hole analytically.

	The trace anomaly equation (\ref{i}) is a direct consequence
of the breaking of Weyl symmetry in the definition of the functional
measure for the conformal matter fields,
\be
    || \delta f_i ||^2 = \int d^2 x \sqrt{-g} \delta f_i \delta f_i
\>;
\label{iii}
\ee
the above definition respects diffeomorphism invariance of the
classical theory but sacrifices Weyl symmetry.
The former symmetry leads to the standard covariant conservation of
the quantum energy-momentum tensor
\be
	\nabla_\mu < {T^f}^{\mu\nu} > = 0 \>,
\label{iv}
\ee
and the effective action capturing the equations (\ref{i}) and (\ref{iv})
is given by the induced gravity action (\ref{ii}).

	Recently, it has been advocated
an alternative definition of the measure (\ref{iii}) that preserves
Weyl invariance \cite{Karakhanyan,Jackiw}.
Diffeomorphism invariance is partially
lost and only area-preserving diffeomorphisms are maintained.
In this note we shall develop further this alternative approach and
explore the relationship between
the diffeomorphism and Weyl invariant schemes in the context of the
semiclassical theory of 2D gravity.

\section{The Weyl-invariant effective action and the Virasoro
anomaly}

	The Weyl-invariant effective action proposed in
\cite{Karakhanyan,Jackiw} can be obtained from the Polyakov
action replacing $g^{\mu\nu}$ by $\sqrt{-g} g^{\mu\nu}$:
\be
S_W = - {N \over 96\pi} \int d^2 x R(\sqrt{-g} g^{\mu\nu} )
				(\sqrt{-g} \square)^{-1}
				R(\sqrt{-g} g^{\mu\nu} )
\>,
\label{v}
\ee
where
$\square = (\sqrt{-g})^{-1} \partial_\mu \sqrt{-g} g^{\mu\nu} \partial_\nu$.
To study the relation between the action (\ref{v}) and (\ref{ii}) in a
simple way it is convenient to introduce an auxiliary field
$\Phi$ verifying the equation
\be
	\square \Phi = R \>.
\label{vi}
\ee
In terms of $g^{\mu\nu}$ and $\Phi$ the Polyakov action can be rewritten
as (see, for instance, \cite{JNavarro})
\be
	S_P = - {N \over 96\pi} \int d^2 x \sqrt{-g}
		( -\Phi \square \Phi + 2 R \Phi )
\>.
\label{vii}
\ee
The equation (\ref{vi}) follows from the action (\ref{vii}) and
inserting (\ref{vi}) into (\ref{vii}) we recover (\ref{ii}).
The metric-dependent Weyl transformation
$g^{\mu\nu} \rightarrow \sqrt{-g} g^{\mu\nu}$ induces the following
transformation for the scalar curvature and the auxiliary field
$\Phi$
\bea
	R &\rightarrow& ( R + \square \log\sqrt{-g} ) \sqrt{-g}
\>, \label{viii}\\
	\Phi &\rightarrow& \Phi + \log\sqrt{-g}
\>. \label{ix}
\eea
Using (\ref{viii}) and (\ref{ix}) it is easy to find that the
Weyl-invariant action $S_W$ is given by
\bea
S_W &=&
	- {N \over 96\pi} \int d^2 x \Bigl[
	\sqrt{-g} (-\Phi \square
	\Phi + 2 R \Phi )
	\nonumber \\
	&&
	\phantom{- {N \over 96\pi} \int d^2 x \Bigl[}
	+ \partial_\mu \bigl( \sqrt{-g}
	( \log\sqrt{-g} {\buildrel\leftrightarrow\over{\partial^\mu}}
	\Phi ) \bigr)
	\nonumber \\
	&&
	\phantom{- {N \over 96\pi} \int d^2 x \Bigl[}
	+ \sqrt{-g} ( \log\sqrt{-g} \square \log\sqrt{-g} +
		2 R \log\sqrt{-g} ) \Bigr]
\>.
\label{x}
\eea
Therefore, $S_W$ and $S_P$ differ, up to total derivative terms, by a
local action
\be
S_W = S_P - {N \over 96\pi} \int d^2 x \sqrt{-g}
	(\log\sqrt{-g} \square \log\sqrt{-g} + 2 R \log\sqrt{-g})
\>.
\label{xi}
\ee

	The energy-momentum tensor $T^W_{\mu\nu} = - {2\pi \over \sqrt{-g}}
{\delta S_W \over \delta g^{\mu\nu}}$ coming from (\ref{xi}) can be
easily computed. It admits the following decomposition
\bea
T^W_{\mu\nu} &=&
	(T^P_{\mu\nu} - {N \over 48}g_{\mu\nu} R) \nonumber \\
	&& - {N \over 48} \left[
		\partial_\mu \log\sqrt{-g} \partial_\nu \log\sqrt{-g} -
		{1\over2} g_{\mu\nu} \partial_\alpha \log\sqrt{-g}
		\partial^\alpha \log\sqrt{-g} \right] \nonumber \\
	&& - {N \over 24} \left[ \nabla_\mu \nabla_\nu \log\sqrt{-g}
		-{1\over2} g_{\mu\nu} \square \log\sqrt{-g} \right]
\>,
\label{xii}
\eea
where $T^P_{\mu\nu}$ is the energy-momentum associated with the
Polyakov effective action
\bea
T^P_{\mu\nu} &=&
	{N \over 48} \left[ \partial_\mu \Phi \partial_\nu \Phi -
	{1\over2} g_{\mu\nu} \partial_\alpha \Phi \partial^\alpha \Phi
	\right] \nonumber \\
	&& - {N \over 24} \left[ \nabla_\mu \nabla_\nu \Phi -
	{1\over2} g_{\mu\nu} \square \Phi \right] \nonumber \\
	&& + {N \over 48} g_{\mu\nu} R
\>.
\label{xiii}
\eea
Note that ${{T^P}^\mu}_\mu = {N \over 24} R$ is consistent with (\ref{i}).

        Due to the breakdown of reparametrization invariance $T^W_{\mu\nu}$
is not covariantly conserved, in general. Instead, we have
\be
\nabla^\mu T^W_{\mu\nu} =
-{N\over48} \left[ \partial_\nu ( R + \square \log\sqrt{-g} ) +
( R + \square \log\sqrt{-g} ) \partial_\nu \log\sqrt{-g} \right]
\>,
\label{xiiia}
\ee
and taking into account that
\be
R + \square \log\sqrt{-g} = {1\over\sqrt{-g}} R(\sqrt{-g} g^{\mu\nu})
\>,
\label{xiiib}
\ee
we can rewrite (\ref{xiiia}) as
\be
\nabla^\mu T^W_{\mu\nu} = -{N\over48} {1\over\sqrt{-g}}
	\partial_\nu R(\sqrt{-g} g^{\mu\nu})
\>.
\label{xiiic}
\ee
This formula has been obtained in \cite{Jackiw} in a different way.
However, in special gauges the stress tensor $T^W_{\mu\nu}$ can be
conserved. For metrics of the form
\be
ds^2 = - e^{2\rho} \left(
dx^+ dx^- +  c \, (x^-)^2  (dx^+)^2 \right)
\>,
\label{xiiid}
\ee
where $c$ is a constant, the r.h.s. of (\ref{xiiic}) vanishes.
In the conformal gauge the conservation equations take the simple form
\be
\nabla^+ T^W_{++} = 0 = \nabla^- T^W_{--}
\>,
\label{xiiie}
\ee
which implies that
\bea
T^W_{++} &=& T^W_{++} (x^+) \>, \label{xiiif} \\
T^W_{--} &=& T^W_{--} (x^-) \>. \label{xiiig}
\eea
Despite of the covariant conservation equations (\ref{xiiie}) the
transformation law of $T^W_{\pm\pm}$ is anomalous.
Under conformal coordinate transformations
$x^\pm \rightarrow y^\pm (x^\pm)$,
$T^W_{\mu\nu}$ transforms as
\be
	T^W_{x^\pm x^\pm} =
	\left( {dy^\pm \over dx^\pm} \right)^2 T^W_{y^\pm y^\pm} -
	{N \over 24} \left\{ y^\pm, x^\pm \right\}
\>,
\label{xiv}
\ee
where
\be
	\left\{ y, x \right\} = { {\partial^3 y \over \partial x^3}
		\over {\partial y \over \partial x} } -
		{3\over2} { \left( {\partial^2 y \over \partial x^2}
		\right)^2 \over \left( {\partial y \over \partial x}
		\right)^2 }
\label{xv}
\ee
is the Schwartzian derivative. Note that the expression (\ref{xiv}) coincides
with the well-known transformation law of the normal-ordered
energy-momentum tensor of a conformal field theory.
Therefore we can conclude that the local counterterm in (\ref{xi})
converts the trace anomaly into the Virasoro anomaly.

\section{Semiclassical 2D black holes. Field redefinitions and covariance}

	The action of the CGHS model is \cite{CGHS}
\be
S = {1\over2\pi} \int d^2 x \sqrt{-g}
	\left[ e^{-2\phi} (R + 4 (\nabla\phi)^2 + 4\lambda^2 ) -
	{1\over2} \sum_{i=1}^N (\nabla f_i)^2 \right]
\>.
\label{xvi}
\ee
The classical solutions describe the formation of a black hole
by gravitational collapse.
In conformal gauge, $ds^2 = -e^{2\rho} dx^+ dx^-$ ($x^\pm = x^0 \pm x^1$),
and in Kruskal coordinates, $\rho=\phi$,
the black hole formation from the vacuum by left moving incoming matter
$f_i= f_i^+(x^+)$ is described by the solution
\be
e^{-2\phi} = e^{-2\rho} = -\lambda^2 x^+ (x^- + {1\over\lambda^2} P(x^+))
	+ {M(x^+) \over \lambda}
\>,
\label{xvii}
\ee
where
\bea
M(x^+) &=& \lambda \int^{x^+}_0 d{\tilde x}^+ {\tilde x}^+ T^f_{++}
\>, \label{xviii}\\
P(x^+) &=& \int^{x^+}_0 d{\tilde x}^+ T^f_{++}
\>, \label{xix}
\eea
and
\be
T^f_{++} = {1\over2} \sum_{i=1}^N ( \partial_+ f^+_i )^2
\>.
\label{xx}
\ee
This solution corresponds to a black hole of mass $M =
M(x^+\rightarrow\infty)$ with an event horizon located at $x^- =
-{1\over\lambda^2}P(x^+\rightarrow\infty)$.
In the semiclassical approximation the Hawking radiation at future null
infinity can be obtained from the trace anomaly \cite{CGHS}, and
the back-reaction is incorporated \cite{CGHS} by adding to the classical
action $S$ the Polyakov term (\ref{ii}).

Let us consider an $f$ shock wave travelling in the $x^-$-direction,
described by the stress tensor
\be
{1\over2} \partial_+ f \partial_+ f = a \; \delta(x^+ - x^+_0)
\> .
\label{xxix}
\ee
For $x^+<x^+_0$ the classical solution is the linear dilaton vacuum (LDV).
In this region the natural coordinate system is the Minkowskian one. We
assume, as boundary condition of the gravitational collapse, that the
outgoing energy flux measured by the Minkowskian observer $\sigma^\pm$
vanishes,
\be
T^W_{\sigma^-\sigma^-} = 0
\>.
\label{xxx}
\ee
After the collapse, $x^+>x^+_0$, the
classical solution describes a black hole of mass $M = a x^+_0 \lambda$ with
a horizon at $x^- = -{a\over\lambda^2} (=-{P\over\lambda^2})$. But
then the natural coordinates are the Schwarzschild type coordinates
${\tilde{\sigma}}^\pm$. The coordinate transformation
\bea
{\tilde{\sigma}}^+ &=& \sigma^+ \>, \label{xxxi} \\
{\tilde{\sigma}}^- &=& -{1\over\lambda} \log\left( e^{-\lambda\sigma^-} -
{a\over\lambda} \right) \>, \label{xxxii}
\eea
allows to evaluate the energy flux measured by the Schwarzschild-type observer
(${\tilde{\sigma}}^\pm$). It is given by
\bea
T^W_{{\tilde{\sigma}}^- {\tilde{\sigma}}^-}
	&=&
- {N\over24} \left\{ \sigma^-, {\tilde{\sigma}}^-
\right\} \>,  \label{xxxiii} \\
T^W_{{\tilde{\sigma}}^+{\tilde{\sigma}}^+} &=& 0 \>, \label{xxxiv}
\eea
and hence
\be
	T^W_{{\tilde\sigma}^-{\tilde\sigma}^-} =
	{\lambda^2 N\over 48}
	\left( 1 - {1\over(1+\frac{1}{\lambda} P e^{\lambda\tilde{\sigma}^-})^2}
	\right)
\>,
\label{xxxv}
\ee
in agreement with the result predicted by the trace anomaly.
The reason of this is that the difference between $T^W_{\mu\nu}$ and
$T^P_{\mu\nu}$ vanishes at infinity in asymptotically flat coordinates.
So both $T^W$ and $T^P$ yield to the same flux of Hawking radiation.

\bigskip

Let us briefly analyze a more involved model: spherically symmetric gravity
coupled to 2D conformal matter.
After appropriate reduction this model is described by the two-dimensional
action (see, for instance, \cite{Strominger})
\be
S= {1\over2\pi} \int d^2 x \sqrt{-g} \left[
e^{-2\phi} (R + 2 (\nabla\phi)^2 + 2 \lambda^2 e^{2\phi}) -
{1\over2}\sum_{i=1}^N (\nabla f_i)^2 \right]
\>,
\label{xxxva}
\ee
where the 4D spherically symmetric metric is related to the 2D metric and
dilaton fields by ${}^{(4)}ds^2 = ds^2 + e^{-2\phi} / \lambda^2 d\Omega^2$.
This model is classically soluble and the solutions are
the Vaidya space-times.
If we consider the collapse of a null shell of matter, the stress tensor and
two-dimensional metric take the form
\bea
T^f_{vv} &=& m \delta(v-v_0) \>, \nonumber \\
ds^2 &=& - \left( 1 - {2m\theta(v-v_0) \over r} \right) dv^2 + 2 dr dv
\>.
\label{xxxvb}
\eea
In conformal coordinates matching the discontinuity across $v=v_0$ we obtain
for the metric
\be
ds^2 =
-\left[
\theta (v_0-v) + \theta (v-v_0)
\left( 1 -{2m\over r} \right) \left( 1 -{4m\over v_0-u} \right)^{-1}
\right]
dv du
\>,
\label{xxxvc}
\ee
where $r$ is defined implicitly by
\be
u - 4m \log\left( {v_0-u\over 4m} -1 \right) - v = - 2 \left( r +
2m\log\left( {r\over2m} -1 \right) \right)
\>,
\label{xxxvd}
\ee
and for the dilaton
\be
{e^{-\phi}\over\lambda} = {v-u \over 2} \theta(v_0-v) + r \theta(v-v_0)
\>.
\label{xxxve}
\ee
The coordinates $(v,u)$ are Minkowskian inside the shell ($v<v_0$),
whereas the asymptotically flat conformal coordinates
$(\tilde{v},\tilde{u})$ are given by the relations
\bea
\tilde{v} &=& v \nonumber \>, \\
\tilde{u} &=& u - 4 m \log \left( {v_0 - u \over 4m} - 1 \right) \>.
\label{xxxvf}
\eea
Therefore, assuming that $T^W_{uu}=0$, the stress tensor evaluated by the
Schwarzschild observer is given by
\be
T^W_{\tilde{u}\tilde{u}} = - {m (u - v_0 + 3m) \over 3 (u - v_0)^4}
\>.
\label{xxxvg}
\ee
As the horizon is approached $u \rightarrow v_0-4m$,
$T^W_{\tilde{u}\tilde{u}}$ builds up to the value $(3\cdot 2^8 m^2)^{-1}$,
that corresponds to the Hawking temperature $T={1\over8\pi m}$.

\bigskip

Our aim now is to study the semiclassical back-reaction, for the CGHS model,
defined by the Weyl-invariant
effective action (\ref{xi}), in the large $N$ limit.
Due to Weyl invariance the $\rho$, $\phi$ classical equations are unmodified,
\bea
e^{-2\phi} \left(
2 \partial_+\partial_-\phi - 4 \partial_+\phi \partial_-\phi -
\lambda^2 e^{2\rho}
\right) &=& 0 \>, \label{xxvi} \\
e^{-2(\phi+\rho)} \left(
-4 \partial_+\partial_-\phi + 4 \partial_+\phi \partial_-\phi +
2 \partial_+\partial_-\rho + \lambda^2 e^{2\rho}
\right) &=& 0 \>.
\label{xxvii}
\eea
However, the constraint equations are modified according to
\be
e^{-2\phi} \left(
4 \partial_\pm \rho \partial_\pm\phi - 2 \partial_\pm^2 \phi
\right)
+ {1\over2} \sum_{i=1}^N \left( \partial_\pm f_i \right)^2 +
T^W_{\pm\pm} = 0
\> .
\label{xxviii}
\ee
The components $T^W_{\pm\pm}$ of the non-local effective stress tensor are
$\rho$-independent and have to be adjusted by boundary conditions.

At this point we should note that, although the Virasoro anomaly of
$T^W_{\pm\pm}$ accounts for the Hawking radiation, it destroys the
covariance of the one-loop equations (\ref{xxvi}-\ref{xxviii}).
Moreover, if
we select a particular coordinate system ($\sigma^\pm$ before the collapse
and $\tilde{\sigma}^\pm$ after it) the solution is not continuous at
$x^+=x^+_0$; and coordinates that match continuously $\sigma^\pm$ and
$\tilde{\sigma}^\pm$ do no adjust to the gauge (\ref{xiiid}).
Then, how can we define, in this scheme, a consistent semiclassical theory?.
To gain insights, let us first analyse the question of the vacuum stability.

In conformal gauge, the vacuum solution to equations
(\ref{xxvi}-\ref{xxviii}) is
\bea
e^{-2\rho} &=& e^{-\omega} e^{-2\phi} \>, \nonumber \\
e^{-2\phi} &=& u - h_+ h_- \>,
\label{xxxvi}
\eea
where $\omega = \omega_+ (y^+) + \omega_- (y^-)$, $u = u_+ + u_-$,
\bea
u_\pm &=& -\int^{y^\pm} e^{\omega_\pm} \int e^{-\omega_\pm}
T^W_{\pm\pm} \>, \nonumber \\
h_\pm &=& \lambda \int^{y^\pm} e^{\omega_\pm} \>.
\label{xxxvii}
\eea
According to (\ref{xiv}), $T^W_{y^\pm y^\pm}$ is given by the Schwartzian
derivative with respect to the Minkowskian coordinates $\sigma^\pm$.
Hence the general solution to the semiclassical equations in vacuum
is found to be
\bea
e^{-2\rho} &=&
e^{-2(\rho_{cl} - \phi_{cl})} \Bigl[
e^{-2\phi_{cl}} + {N\over24}(2\rho_{cl}-\phi_{cl}) \nonumber \\
&&
\phantom{e^{-2(\rho_{cl} - \phi_{cl})} }
+ {N\over12} \int^{y^+} e^{2(\rho_{cl}-\phi_{cl})} \int
e^{-2(\rho_{cl}-\phi_{cl})} [ \partial_{y^+} (\rho_{cl} - \phi_{cl}) ]^2
\nonumber \\
&&
\phantom{e^{-2(\rho_{cl} - \phi_{cl})} }
+ {N\over12} \int^{y^-} e^{2(\rho_{cl}-\phi_{cl})} \int
e^{-2(\rho_{cl}-\phi_{cl})} [ \partial_{y^-} (\rho_{cl} - \phi_{cl}) ]^2
\Bigr]
\>,
\label{xxxviii}
\eea
where $\rho_{cl}$ and $\phi_{cl}$ represent the classical vacuum solutions
\bea
\rho_{cl} &=& {1\over2} \log {dy^+ \over d\sigma^+} {dy^- \over d\sigma^-}
\>, \nonumber \\
\phi_{cl} &=& -{\lambda\over2} \left( \sigma^+(y^+) - \sigma^-(y^-) \right)
\>.
\label{xxxix}
\eea
In the light of the above expressions we can conclude that, due to the
term $T^W_{\pm\pm} (y^\pm)$, the solutions do no transform
covariantly, thus producing some sort of vacuum instability that makes
problematic a Weyl-invariant semiclassical theory.

A way to construct a sensible semiclassical theory is suggested by the
non-covariant vacuum solutions themselves (\ref{xxxviii}).
The non-local field redefinition $\rho \rightarrow \hat{\rho}$, $\phi
\rightarrow \hat{\phi}$, defined by
\bea
\rho-\phi &=& \hat{\rho} - \hat{\phi} \nonumber \\
e^{-2\phi} &=&
\Bigl[
e^{-2\hat{\phi}} +
{N\over24}(2\hat{\rho}-\hat{\phi}) \nonumber \\
&&
\phantom{\Bigl[}
+ {N\over12} \int^{y^+} e^{2(\hat{\rho}-\hat{\phi})} \int
e^{-2(\hat{\rho}-\hat{\phi})} [ \partial_{y^+} (\hat{\rho} - \hat{\phi}) ]^2
\nonumber \\
&&
\phantom{\Bigl[}
+ {N\over12} \int^{y^-} e^{2(\hat{\rho}-\hat{\phi})} \int
e^{-2(\hat{\rho}-\hat{\phi})} [ \partial_{y^-} (\hat{\rho} - \hat{\phi}) ]^2
\Bigr]
\>,
\label{xxxx}
\eea
recovers the classical vacuum solution as a solution of the semiclassical
vacuum equations:
$\hat{\rho} = \rho_{cl}$, $\hat{\phi} = \phi_{cl}$, and reestablishes the
general covariance as well.

After the redefinitions (\ref{xxxx}), the constraint equations
(\ref{xxviii}) become
\bea
\left(
e^{-2\hat\phi} + {N\over48}
\right)
\left (
4 \partial_\pm \hat\rho \partial_\pm \hat\phi - 2 \partial^2_\pm \hat\phi
\right)
- {N\over12}
\left(
\partial_\pm \hat\rho \partial_\pm \hat\rho - \partial^2_\pm \hat\rho
\right) && \nonumber \\
+ {1\over2} \partial_\pm f_i \partial_\pm f_i
+ T^W_{\pm\pm}
&=& 0
\>,
\label{xxxxii}
\eea
and the equations (\ref{xxvi}-\ref{xxvii}) can be rewritten as
\bea
\partial_+ \partial_- \left( \hat\rho - \hat\phi \right) &=& 0 \>, \\
\partial_+ \partial_- \left( e^{-2\hat\phi} + {N\over24} \hat\rho \right)
&=& -\lambda^2 e^{2(\hat\rho - \hat\phi)}
\> .
\eea
Remarkably these equations
turn out to be equivalent to the equations of motion of the RST model
\cite{RST}.
Note that in Kruskal coordinates $\hat\rho = \hat\phi$ the transformation
(\ref{xxxx}) is local,
\be
e^{-2\rho} = e^{-2\hat{\rho}} +  {N\over24} \hat{\rho}
\>,
\label{xxxxi}
\ee
and coincides with the field redefinitions of \cite{RST} that maps
(\ref{xxxxii}) into a Liouville theory (see also \cite{Bilal}),
where the original variable $e^{-2\phi}$ is the analogue of the
$\sqrt{{N\over12}}\Omega$ field.

This intriguing relation between the RST model and the Weyl-invariant
one-loop CGHS model offers a new explanation of the solubility of the former.
In the Weyl-invariant scheme the semiclassical equations can be solved in a
straightforward way by replacing the classical stress tensor $T^f_{\pm\pm}$
by $T^f_{\pm\pm} + T^W_{\pm\pm}$ in the expression of the classical
solutions.
In other words, the classical solubility of the model together with the
field redefinitions that ensure the vacuum stability allows to solve
a related covariant semiclassical model.

\section{Final comments}

We have exhibit in a simple way the relationship between diffeomorphism and
Weyl-invariant effective actions of conformal matter coupled to 2D gravity.
They differ in a local term that converts the trace anomaly of the effective
stress tensor into the Virasoro anomaly thus yielding to the same flux of
Hawking radiation.
In special gauges the semiclassical equations of the Weyl-invariant
effective action are consistent with Bianchi identities
although the solutions do not transform covariantly.
A consequence of this is the impossibility of defining a covariant vacuum
solution.
To reestablish the LDV as the semiclassical vacuum solution of the CGHS model,
we are forced
to perform specific, non-local field redefinitions of the metric and
dilaton fields.
The field redefinitions yields to new covariant semiclassical equations,
which turn out to be the equations of the RST model.
The solubility of the classical model and the Weyl invariance of the
semiclassical correction implies exact solubility of the related
semiclassical covariant model.
This mechanism can be applied to solve the semiclassical theory of more
realistic models, as
4D-spherically symmetric gravity coupled to 2D conformal matter
\cite{inprogress}.
Moreover, the insights gained in unravelling the relationship between the
Weyl and diffeomorphism invariant schemes could be of great interest for the
non-perturbative canonical approach.
There are some arguments \cite{Jackiw}
indicating that the Weyl invariant scheme reflects more appropriately
the exact canonical quantization.

\section*{Acknowledgements}

	M. Navarro  acknowledges to the {\it MEC} for a Postdoctoral
fellowship.
C. F. Talavera is grateful to the {\it Generalitat Valenciana}
for a FPI grant.

\end{document}